\documentstyle[preprint,aps,epsfig]{revtex}
\begin{document}

\title{Hermitian quark mass matrices with four texture zeros}

\author{Shao-Hsuan Chiu$^{1}$\thanks{chiu@physics.purdue.edu},
T. K. Kuo$^{1}$\thanks{tkkuo@physics.purdue.edu}, and 
Guo-Hong Wu$^{2}$\thanks{wu@dirac.uoregon.edu}} 

\address{$^{1}$Department of Physics, Purdue University, West Lafayette, IN 47907}

\address{$^{2}$Institute of Theoretical Science, University of Oregon,
Eugene, OR 97403}


\maketitle


\begin{abstract}

We provide a complete and systematic analysis of hermitian, 
hierarchical  quark mass matrices 
with four texture zeros.
Using triangular mass matrices,
each pattern of texture zeros is readily shown to lead to a definite relation 
between the CKM parameters and the quark masses. 
Nineteen  pairs are found to be consistent with present data,
 and one other is marginally acceptable.
In particular, no parallel structure between the up and down mass
matrices is found to be favorable with data.

\end{abstract}


\pagenumbering{arabic}

\section{Introduction}

The standard model is known to have a large number of parameters.  However,
most of them are contained in the pair of quark mass matrices,
$M_{U}$ and $M_{D}$.  
Of the thirty-six parameters in these two complex matrices, 
only ten [six quark masses, three Cabibbo-Kobayashi-Maskawa (CKM) 
angles and a CP phase] are physical.
This redundancy arises from unphysical, right-handed (RH) rotations in both
the $U$- and $D$-sectors, in addition to common left-handed (LH) rotations,
which cancel out in the CKM matrix,
$V^{\mbox{\scriptsize CKM}}=V_{U}^{\dag}V_{D}$, 
where $V_{U}^{\dag}M_{U}U_{U}=M_{U}^{\mbox{\scriptsize diag}}$
and $V_{D}^{\dag}M_{D}U_{D}=M_{D}^{\mbox{\scriptsize diag}}$.  
To eliminate these unphysical     
degrees of freedom, it was pointed out earlier~\cite{kmw1:99} that 
RH rotations can be used to reduce both $M_{U}$ and $M_{D}$ 
to the upper triangular form,
which in the hierarchical basis exhibits most clearly the quark masses
and left-handed (LH) rotation angles,
\begin{equation}
M_{U}U_{U}^{R}=M_{U}^{t}=
   \left(\begin{array}{ccc}
    X & X & X \\
    0 & X & X \\
    0 & 0 & X \\
    \end{array}
    \right),
\end{equation}

\begin{equation}
M_{D}U_{D}^{R}=M_{D}^{t}=
   \left(\begin{array}{ccc}
    X & X & X \\
    0 & X & X \\
    0 & 0 & X \\
    \end{array}
    \right). 
\end{equation}

The matrices $U_{U}^{R}$ and $U_{D}^{R}$ can be constructed explicitly
as $R_{13}R_{23}R_{12}$, such that $R_{13}$ eliminates the (3,1) element of
$M$, etc.  Another way to obtain the triangular matrix elements directly
is by a geometric argument. Let us write a general (real) matrix $M$
in the form:
\begin{equation}
 M=
\left(\begin{array}{ccc}
    a_{1} & a_{2} & a_{3} \\
    b_{1} & b_{2} & b_{3} \\
   c_{1}  & c_{2} & c_{3} \\
    \end{array}
    \right).
\end{equation}
We may regard the three rows as components of the vectors
$\vec{a}, \vec{b}$ and $\vec{c}$.  Then the transformation
into the triangular form amounts to a rotation into a new coordinate
system where the axes are aligned in the directions of the unit vectors:
$(\vec{b} \times \vec{c})/|\vec{b} \times \vec{c}|$, 
$\vec{c} \times (\vec{b} \times \vec{c})/(|\vec{c}| \cdot |\vec{b} 
\times \vec{c}|)$,
and $\vec{c}/|\vec{c}|$, respectively.  It follows that

\begin{equation}
           M^{t}= 
\left(\begin{array}{ccc}
 \frac{\vec{a}\cdot (\vec{b} \times \vec{c})}{\mid \vec{b} \times \vec{c} \mid} &

\frac{(\vec{a} \times \vec{c}) \cdot (\vec{b} \times \vec{c})}
{\mid \vec{c} \mid \cdot \mid \vec{b} \times \vec{c} \mid} & 
\frac{\vec{a} \cdot \vec{c}}{\mid \vec{c} \mid} \\
    0 &
\frac{\mid \vec{b} \times \vec{c} \mid}{\mid\vec{c}\mid} & 
\frac{\vec{b} \cdot \vec{c}}{\mid \vec{c} \mid} \\

  0     & 0 & \mid \vec{c} \mid \\
    \end{array}
    \right). 
\end{equation}
For complex vectors $\vec{a}, \vec{b}$ and $\vec{c}$, each transforms as 
a \textbf{3} under RH SU(3) rotations.  
From $[\mathbf{3} \times \mathbf{3}]_{anti} \sim \overline{\mathbf{3}}$,
 $[\mathbf{3} \times \mathbf{3} \times \mathbf{3}]_{anti} \sim \mathbf{1}$, 
etc., we readily find:
 \begin{equation}
           M^{t}= 
\left(\begin{array}{ccc}
    \frac{\vec{a}\cdot (\vec{b} \times \vec{c})}{\mid \vec{b} \times \vec{c} \mid} &

\frac{(\vec{a} \times \vec{c}) \cdot (\vec{b^{*}} \times \vec{c^{*}})}
{\mid \vec{c} \mid \cdot \mid \vec{b} \times \vec{c} \mid} & 
\frac{\vec{a} \cdot \vec{c^{*}}}{\mid \vec{c} \mid} \\
    0 &
\frac{\mid \vec{b} \times \vec{c} \mid}{\mid\vec{c}\mid} &
\frac{\vec{b} \cdot \vec{c^{*}}}{\mid \vec{c} \mid} \\

  0     & 0 & \mid \vec{c} \mid \\
    \end{array}
    \right).
\end{equation}
Note that this construction is unique up to a diagonal phase matrix
multiplying from the right:
$M^{t} \rightarrow M^{t}P$, $P=$ diagonal phase matrix.

After we have transformed $M_{U}$ and $M_{D}$ to the upper triangular form,
we may reduce them further by common LH rotations on both.
Since there are three degrees of freedom in these rotations, we can use them to
generate three additional zeros in the pair.  Note that this result is only
approximately correct.  The LH rotations will generate small elements in the
lower-left part of the matrices, which can be removed by even smaller RH
rotations.  Thus, the process actually consists of a sequence of rotations
$R_{i}$, which can be schematically expressed as 
$\cdot \cdot R_{3}((\frac{m_{1}}{m_{2}})^{2}V_{12},
(\frac{m_{2}}{m_{3}})^{2}V_{23}, (\frac{m_{1}}{m_{3}})^{2}V_{13}) 
R_{1}(V_{12}, V_{23}, V_{13}) M R_{2}(\frac{m_{1}}{m_{2}}V_{12},
\frac{m_{2}}{m_{3}}V_{23}, \frac{m_{1}}{m_{3}}V_{13})\cdot \cdot $.

The above argument provides an algorithm to reduce any pair of
$M_{U}$ and $M_{D}$ into upper triangular forms, with the
minimal nine non-vanishing
elements between the two matrices.  As was shown earlier, the most interesting
property of this reduction is that to a good approximation, all of the
remaining nine non-vanishing matrix elements in $M^{U}$ and $M^{D}$ are
physical and are simple products of a quark mass and some CKM matrix
elements.  This method can thus provide us with a powerful tool to
assess the viability of any proposed mass matrices.  It was applied 
to hermitian mass matrices with five texture zeros, and 
a unique pair is identified as most favorable to present data~\cite{kmw2:99}.  

In this paper, we will apply the same technique to the study of hermitian 
mass matrices with four texture zeros, which generally entail 
one testable relation for each pair. 
The study of 4-texture-zero hermitian mass matrices has been a
subject of interest recently [3-8]. 
Attentions have been focused on specific models or structures, 
which can lead to predictive values for the CKM matrix elements. 
However, due to the complexity involved in the analysis, no complete
investigation has been performed so far.
Using the triangular basis can greatly reduce the amount of work 
involved in analyzing the viability of 4-zero textures. 
With this efficient tool, we are able to provide the first complete 
study of all viable four-zero texture pairs that 
exhibit hierarchical
structure.
The results are presented in Tables I through IV.

\section{Hermitian Mass Matrices and Texture Zeros}

We start with quark mass matrices in the triangular form.
   Using the mass relations $m_u:m_c:m_t \sim \lambda^8:\lambda^4:1$
and $m_d:m_s:m_b \sim \lambda^4:\lambda^2:1$,
and the CKM elements
$V_{us} = \lambda$, $V_{cb} \sim \lambda^2$,
$V_{ub} \sim \lambda^4$, and $V_{td} \sim \lambda^3$,  the
properly normalized Yukawa matrices for $U$ and $D$
can be put into the most general triangular form,
\begin{equation} \label{eq:triUD}
T^U = \left(\begin{array}{ccc}
       a_U \lambda^8 & b_U \lambda^6 & c_U \lambda^4 \\
          0          & d_U \lambda^4 & e_U \lambda^2 \\
          0          &    0          &  1
           \end{array}\right)\; ,
\;\;\;\;\;\;\;\;\;\;
T^D = \left(\begin{array}{ccc}
       a_D \lambda^4 & b_D \lambda^3 & c_D \lambda^3 \\
          0          & d_D \lambda^2 & e_D \lambda^2 \\
          0          &    0          &  1
           \end{array}\right)\; .
\end{equation}
Here, all of the coefficients ($a, b,\cdot \cdot \cdot$) are assumed to be 
of order unity or less. Without loss of generality, we also take
the diagonal parameters (i.e. $a_U$, $a_D$, $d_U$, $d_D$) to be real
throughout this paper.
  We offer the following remarks.
\begin{enumerate}
  \item  The hierarchical
structures of $T^{U}$ and $T^{D}$ are manifest.  
As was shown earlier\cite{kmw1:99,kmw2:99}, 
we can diagonalize them approximately with only a LH rotation,
$M_{U,D}^{\mbox{\scriptsize diag}} \simeq V_{U,D}^{\dag} T^{U,D}$,
with the matrix elements
$(V_{U,D})_{ij} \simeq T_{ij}^{U,D}/T_{jj}^{U,D}$ ($i<j$).
  \item  The
diagonal elements of $T^{U}$ and $T^{D}$ are essentially the quark masses.
The CKM matrix elements are simply given by 
$V_{ij}^{\mbox{\scriptsize CKM}}
 \sim (V_{D})_{ij}-(V_{U})_{ij}$, $\mbox{ }$ ($i<j$). 
These simple relations are lost when we go to other bases, including
the hermitian basis.
Triangular matrices thus stand out as the unique basis whose matrix
elements correspond directly to quark masses and LH rotation angles. 
This feature makes the triangular form especially useful for analyzing
the texture of quark mass matrices.
\item  To avoid fine tuning in generating the CKM mixings,
 naturalness criteria~\cite{PW} has been imposed when writing 
the above triangular matrices.
This simply implies that, for the LH rotation angles, 
$|V_{U,D}|_{ij} \leq  |V_{ij}^{\mbox{\scriptsize CKM}}| 
\times {\cal O} (1) $.  
Note that this condition can always be
implemented by applying a common LH rotation on $T^{U}$ and $T^{D}$. 
\item We can put $T^{U}$ and $T^{D}$ in 
 the minimal parameter basis (m.p.b.),
with only three nonzero off-diagonal elements between them.
This is achieved from Eq.~(\ref{eq:triUD}) by common LH rotations to
generate three zeros.  
For example, the (1,3) element of $T^{D}$ can be
set to zero by a common LH rotation $R_{13}(-c_{D}\lambda^{3})$.
This method produces ten pairs of triangular matrices, as listed
in Table~I of Ref.\cite{kmw2:99}.  
The elements of each pair consist of a simple product of 
quark mass and CKM elements. Nothing is unphysical in the m.p.b..
As a result, the viability of a given texture can be readily
tested by turning it into one of the ten.
\end{enumerate}
 
  Turning now to hermitian matrices, the consequences of
their texture zeros  can be easily understood in terms of the
triangular parameters of Eq.~(\ref{eq:triUD}).
For this purpose, one can start from Eq.~(\ref{eq:triUD}) and, 
by a RH rotation, generate the corresponding hermitian form,
\begin{eqnarray} \label{eq:HU}
Y^U  & = & \left(\begin{array}{ccc}
       \left(a_U +  c_U c_U^* +
               \frac{b_Ub_U^*}{d_U}  \right) \lambda^8 &
       (b_U + c_U e_U^*)  \lambda^6 & c_U \lambda^4 \\
       (b_U^* + c_U^* e_U)  \lambda^6  &
       (d_U + e_U e_U^*) \lambda^4 & e_U \lambda^2 \\
          c_U^* \lambda^4  & e_U^*\lambda^2   &  1
           \end{array}\right) \times (1+{\cal O}(\lambda^4)) \; ,
\end{eqnarray}
\begin{eqnarray} \label{eq:HD}
Y^D & = & \left(\begin{array}{ccc}
      \left(a_D +  \frac{b_Db_D^*}{d_D} \right) \lambda^4 &
       b_D \lambda^3 & c_D \lambda^3 \\
       b_D^* \lambda^3  & d_D \lambda^2 & e_D \lambda^2 \\
        c_D^* \lambda^3  &    e_D^* \lambda^2     &  1
           \end{array}\right) \times (1+{\cal O}(\lambda^2)) \ .
\end{eqnarray}
We immediately see that whereas off-diagonal hermitian zeros
(e.g. $Y^U_{13}$, $Y^U_{23}$ or $Y^D_{ij}$ ($i\neq j$))
have a one-to-one correspondence to zeros in the  triangular form,
the diagonal hermitian zeros (i.e. $Y^U_{11}$, $Y^U_{22}$, and $Y^D_{11}$)
entail definite relations between diagonal and
off-diagonal triangular parameters (i.e. quark masses and LH rotation
angles).  Note that these simple relations between the hermitian and
triangular forms are approximate, and follow from the hierarchical
structure in Eq.~(\ref{eq:triUD}).

We are thus led to the following procedure in analyzing hermitian
texture zeros. 
First, we obtain all possible textures with certain number of zeros
for $Y^U$ and $Y^D$ by referring to Eq.~(\ref{eq:HU}) and
Eq.~(\ref{eq:HD}), as listed in Tables I and III.
 Then we can list all the 4-texture-zero pairs.
Second, each pair is transformed into  one of the ten triangular
forms in the m.p.b., using common LH rotations when necessary.
 Possible relations between quark masses and mixing can then be read-off
for each pair by referring to Table I of Ref.~\cite{kmw2:99}.
  Finally, we arrive at all the viable textures after confronting
the prediction of each pair with data.
In the  next section, we apply this procedure  to the analysis of
hermitian matrices with  four texture zeros.

\section{Analyses of Matrices with Four Texture Zeros}

The study of viable 4-texture-zero pairs is straightforward 
by following the procedure outlined above. 
 For convenience in our analysis and presentation, 
 we categorize the mass matrix pairs
$(M_{U},M_{D})$ according to whether or not
 the $(1,1)$ matrix element is zero: \\
1) $M_{U}^{11}=0, M_{D}^{11}=0$, \\
2) $M_{U}^{11}\neq 0, M_{D}^{11}=0$, \\
3) $M_{U}^{11}=0, M_{D}^{11}\neq 0$, \\
4) $M_{U}^{11}\neq 0, M_{D}^{11}\neq 0$. \\ 
The results are presented in Tables~I-IV, which contain all the 
viable  4-texture-zero pairs.
It is seen from the Tables that almost all of the viable textures 
have $M_{D}^{11}=0$.

The first category involves eight types of hermitian matrices, which are
listed in Table~I along with their corresponding triangular forms.
These hermitian matrices contain 1, 2 and 3 texture zeros for
$M_U$ and/or  $M_D$.   Their corresponding triangular matrices can be simply
obtained from Eqs.~(\ref{eq:HU}) and (\ref{eq:HD}).
 They are then paired up to form possible patterns for the quark mass 
matrices. 
Note that $M_{2}, M_{3}, M_{6}, M_{7}$ 
and $M_{8}$ are the ones appearing in the analysis of five 
texture-zero matrices~\cite{RRR:93,ks,kmw2:99}. 
One notable feature about the pairing is that $M_U$ and $M_D$ allow
different texture zeros, as can be seen from 
Eqs.~(\ref{eq:HU}) and (\ref{eq:HD}).
For example, whereas $M_D^{22} \neq 0$ 
(unless we give up  naturalness),  $M_U^{22}=0$ is allowed due to the
much larger mass hierarchy in the up-quark sector.
As a second example, $M_U^{11}=M_U^{22}=0$ is allowed 
but the same relation is not valid for
$M_D$ again because of the different mass hierarchies.
As another tip for the pairing,
a quick check on the CKM matrix element 
can be useful in the screening of possible candidates.
For instance, the $(M_{4},M_{4})$ pair is out because both their (2,3) 
elements are zero,  which in turns gives $V_{cb}\simeq 0$. 
The same can not be said about hermitian pairs with vanishing 
(1,3) elements.
Here,  $V_{ub}$ and $V_{td}$ may or
  may not vanish since they are of higher order in $\lambda$, and 
may be induced from a combined
  $R_{12}$ and $R_{23}$ rotations.
In this latter case, a further investigation is required. 

The viability of each $(M_{U},M_{D})$ pair can be assessed by converting
them, through common LH rotations if necessary, 
into one of the ten triangular pairs in the m.p.b. as  listed in Table I of
Ref.~\cite{kmw2:99}.  Testable relations can then be obtained.  To
illustrate the method, we first study the $(M_{8},M_{1})$ pair in detail. 
The triangular forms for $M_{8}$ and $M_{1}$ are given by:

\begin{equation}
 M_{8} \rightarrow M_{U}^{t} \simeq
\left(\begin{array}{ccc}
    a_{U}\lambda^{8} & b_{U}\lambda^{6} & 0 \\
    0 & d_{U}\lambda^{4} & e_{U}\lambda^{2} \\
   0  & 0 & 1 \\
    \end{array}
    \right),
\end{equation} 
\begin{equation}
M_{1} \rightarrow M_{D}^{t} \simeq
\left(\begin{array}{ccc}
    a_{D}\lambda^{4} & b_{D}\lambda^{3} & c_{D}\lambda^{3} \\
    0 & d_{D}\lambda^{2} & e_{D}\lambda^{2} \\
   0  & 0 & 1 \\
    \end{array}
    \right).
\end{equation}
Here, $d_{U}=-|e_{U}|^{2}$.  
This ($M_{U}^{t},M_{D}^{t}$) pair has eleven nonzero elements,
and two more zeros are needed to put it into one of the 
ten triangular pairs in the m.p.b..
There are several ways to achieve this. 
For example, a common LH 2-3 rotation with 
$\theta_{23} \simeq -e_{U}\lambda^{2}$ will set the $(2,3)$ element of 
$M_U$ to zero, and a common LH 1-2 rotation with
$\theta_{12} \simeq -\frac{b_{D}}{d_{D}}\lambda$ will 
generate a second zero at the $(1,2)$ position of $M_D$.
In this way, we arrive at  

\begin{equation} \label{eq:MpU}
 M_{U}^{t'}=R_{12}(-\frac{b_{D}}{d_{D}}\lambda)R_{23}(-e_{U}\lambda^{2})M_{U}^{t} 
 \simeq
\left(\begin{array}{ccc}
    a_{U}\lambda^{8} & b_{U}\lambda^{6}-\frac{b_{D}}{d_{D}}d_{U}\lambda^{5} & 0 \\
    0 & d_{U}\lambda^{4} & 0 \\
   0  & 0 & 1 \\
    \end{array}
    \right),
\end{equation}
\begin{equation}  \label{eq:MpD}
M_{D}^{t'}=R_{12}(-\frac{b_{D}}{d_{D}}\lambda)R_{23}(-e_{U}\lambda^{2})M_{D}^{t}
 \simeq
\left(\begin{array}{ccc}
    a_{D}\lambda^{4} & 0 & c_{D}\lambda^{3}-
\frac{b_{D}}{d_{D}}(e_{D}-e_{U})\lambda^{3} \\
    0 & d_{D}\lambda^{2} & (e_{D}-e_{U})\lambda^{2} \\
   0  & 0 & 1 \\
    \end{array}
    \right).
\end{equation}

Identifying $(M_{U}^{t'},M_{D}^{t'})$ with the 5th triangular 
pattern in Table I of Ref.~\cite{kmw2:99} gives, 
\begin{equation} \label{eq:R3a}
  \lambda^{2}b_{U}/d_{U}-\lambda b_{D}/d_{D}=-V_{us}/V_{cs} \; .
\end{equation}
As $M_{U}^{11}=M_{D}^{11}=0$ in the hermitian form, we have
two corresponding relations in terms of triangular parameters:
$a_{U}=-\frac{|b_{U}|^{2}}{d_{U}}$ and $a_{D}=-\frac{|b_{D}|^{2}}{d_{D}}$.
 With these two relations, Eq.~(\ref{eq:R3a}) can be written in the well
known form,
\begin{equation} \label{eq:R3}
\left|\frac{V_{us}}{V_{cs}} \right| \simeq
\left|\sqrt{\frac{m_d}{m_s}} - e^{i\delta} \sqrt{\frac{m_u}{m_c}} \right|,
\end{equation}
where $\delta \equiv \mbox{arg} [b_Ud_D/d_Ub_D]$.
 From  Eqs.~(\ref{eq:MpU}-\ref{eq:MpD}) and  Table I of Ref.~\cite{kmw2:99},
we note that the standard model $CP$-violation depends on additional phases
besides $\delta$, thus leaving $\delta$ a free parameter.
In this sense, Eq.~(\ref{eq:R3}) places a constraint on the mass matrices
in fixing the phase $\delta$, but it is not a prediction in
terms of physical quantities alone.
We conclude that the $(M_{8},M_{1})$ pair is a viable texture  since
Eq.~(\ref{eq:R3}) can be made valid with a properly chosen phase $\delta$.

To assess the viability of any given texture, 
we will need to know about the quark masses and CKM elements.
We use for the quark masses at $m_Z$ 
values  taken from Ref.~\cite{fusaoka:98}.
The CKM matrix elements are taken from \cite{caso:98},
except for $V_{ub}$ and $V_{td}$, for which we use a recent 
update~\cite{parodi,ali}:  
\begin{equation} \label{eq:Vubexp}
  \left|\frac{V_{ub}}{V_{cb}}\right|_{exp} =0.093\pm 0.014 ,
\end{equation}
\begin{equation} \label{eq:Vtdexp}
  0.15<\left|\frac{V_{td}}{V_{ts}}\right|_{exp}<0.24 .
\end{equation}
Whereas Eq.~(\ref{eq:Vubexp}) comes from an average of the LEP and SLD 
measurements, Eq.~(\ref{eq:Vtdexp}) comes from a standard model fit to 
the  electroweak data.
Note that Eq.~(\ref{eq:Vtdexp}) implies $|V_{td}|<0.01$. 

    After a straightforward analysis of all possible texture pairs,
we arrive at the nine viable pairs of hermitian matrices for the
first category. These are listed in Table~II, together with their
testable relations. Note that relations R1 and R2 allow two different
solutions depending on the sign of $d_U$, and that
 some textures share the same quark mass-mixing relation.

  Also listed in Table~II is the particular
pair (pattern 10) which is parallel in structure: $(M_{3},M_{3})$. 
This texture pair has been the center of focus in recent
studies of hermitian 4-texture-zero quark mass 
matrices\cite{nishiura:99,xing:99,randhawa:99,chkareuli:99,branco:99,hall}.
To analyze the predictions of this  pair, we first write down 
the triangular form,

\begin{equation} \label{eq:MtU3}
 M_{3} \rightarrow M_{U}^{t} \simeq
\left(\begin{array}{ccc}
    a_{U}\lambda^{8} & b_{U}\lambda^{6} & 0 \\
    0 & d_{U}\lambda^{4} & e_{U}\lambda^{2} \\
   0  & 0 & 1 \\
    \end{array}
    \right),
\end{equation}
\begin{equation}  \label{eq:MtD3}
M_{3} \rightarrow M_{D}^{t} \simeq
\left(\begin{array}{ccc}
    a_{D}\lambda^{4} & b_{D}\lambda^{3} & 0  \\
    0 & d_{D}\lambda^{2} & e_{D}\lambda^{2} \\
   0  & 0 & 1 \\
    \end{array}
    \right).
\end{equation}
We need one more zero to put the triangular pair
$(M_{U}^{t}, M_{D}^{t})$ in the m.p.b., and this can be attained by
a common LH 2-3 rotation of $\theta_{23} \simeq e_U \lambda^2$
($\theta_{23} \simeq e_D \lambda^2$) to set the $(2,3)$ element of
$M_{U}^{t}$ ($M_{D}^{t}$) to zero.  For example, we can have

\begin{equation} \label{eq:Mt'U3}
 M_{U}^{t'}=R_{23}(-e_{U}\lambda^{2})M_{U}^{t}
 \simeq
\left(\begin{array}{ccc}
    a_{U}\lambda^{8} & b_{U}\lambda^{6} & 0 \\
    0 & d_{U}\lambda^{4} & 0 \\
   0  & 0 & 1 \\
    \end{array}
    \right),
\end{equation}
\begin{equation} \label{eq:Mt'D3}
M_{D}^{t'}=R_{23}(-e_{U}\lambda^{2})M_{D}^{t}
 \simeq
\left(\begin{array}{ccc}
    a_{D}\lambda^{4} & b_D \lambda^3 & 0 \\
    0 & d_{D}\lambda^{2} & (e_{D}-e_{U})\lambda^{2} \\
   0  & 0 & 1 \\
    \end{array}
    \right).
\end{equation}
Note that $(M_{U}^{t'},M_{D}^{t'})$ corresponds to
the 2nd of the ten triangular pairs in the m.p.b. as listed in 
Table I of Ref~\cite{kmw2:99}. A simple comparison between them
gives the following relations for the hermitian $(M_3, M_3)$ texture:
 \begin{equation} \label{eq:VubVtd0}
\frac{V_{ub}}{V_{cb}} \simeq - \frac{b_U}{d_U} \lambda^2 \; ,
\;\;\;\;\;\;\;\;\;\;\;\;
\frac{V^*_{td}}{V_{cb}V_{cs}} \simeq \frac{b_D}{d_D} \lambda \;,
\end{equation}  
and
 \begin{equation} \label{eq:alpha}
 \mbox{arg} \left[ \frac{b_U d_D}{d_U b_D} \right] \simeq \alpha
  \equiv \mbox{arg} \left[- \frac{V_{td} V_{tb}^*}{V_{ud} V_{ub}^*} \right] \; ,
\end{equation}
where $\alpha$ is the $\alpha$-angle of unitarity triangle.
Now the two zeros at the $(1,1)$ position of $(M_3, M_3)$ imply two 
relations among the triangular parameters:
 $|b_U^2|= -a_U d_U$ and $|b_D^2|= -a_D d_D$.
This allows us to rewrite Eq.~(\ref{eq:VubVtd0}) in the form,
 \begin{equation} \label{eq:Vub}
\left|\frac{V_{ub}}{V_{cb}} \right| \simeq 
\sqrt{\frac{m_u}{m_c}} \; ,
\end{equation}
 \begin{equation} \label{eq:Vtd}
\left|\frac{V_{td}}{V_{ts}}\right| \simeq \sqrt{\frac{m_d}{m_s}}  \;.
\end{equation}

 Using the unitarity of the CKM matrix, the two relations 
in Eq.~(\ref{eq:VubVtd0}) can be combined to give
\begin{eqnarray} \label{eq:Vus2}
\frac{V_{us}}{V_{cs}} & = & \frac{V_{ub}}{V_{cb}} +
 \frac{V^*_{td}}{V_{cb}V_{cs}} 
 \simeq  \frac{b_D}{d_D} \lambda - \frac{b_U}{d_U} \lambda^2  \; ,
\end{eqnarray}
where we have assumed ${\rm det} V^{\mbox{\scriptsize CKM}} =1$ without loss of generality.
Using $|b_U^2|= -a_U d_U$, $|b_D^2|= -a_D d_D$, and Eq.~(\ref{eq:alpha}),
Eq.~(\ref{eq:Vus2}) can be cast in the form,
\begin{equation} \label{eq:Vus3}
\left|\frac{V_{us}}{V_{cs}} \right| \simeq
\left|\sqrt{\frac{m_d}{m_s}} - e^{i\alpha} \sqrt{\frac{m_u}{m_c}} \right|.
\end{equation}
Note that Eq.~(\ref{eq:Vus3}) is similar but different from 
Eq.~(\ref{eq:R3}): while the phase in Eq.~(\ref{eq:Vus3}) is in principle
 fixed by the $CP$-violation parameter $\epsilon_K$ to be
$\alpha \sim \pi/2$ \cite{ali,parodi}, the phase $\delta$ in Eq.~(\ref{eq:R3})
can be varied freely as the standard model $CP$ violation depends on
other phases as well.
The relations in Eqs.(\ref{eq:Vub}), (\ref{eq:Vtd}), and
(\ref{eq:Vus3}) were also obtained in 
Ref.~\cite{nishiura:99,xing:99,randhawa:99,chkareuli:99,branco:99,hall}.     

  Whereas Eqs.~(\ref{eq:Vtd}) and ~(\ref{eq:Vus3}) are allowed by
data, the prediction of Eq.~(\ref{eq:Vub}), 
$V_{ub}/V_{cb} = 0.059 \pm 0.006$, is disfavored by the recent 
measurement (see Eq.~(\ref{eq:Vubexp}) ). 
The same conclusion has been reached for three of the five hermitian
pairs in our five-texture-zero analysis\cite{kmw2:99}.
It is interesting and surprising 
 that going to four-texture-zero does not help
with this problem of low $V_{ub}/V_{cb}$.
Furthermore, even varying $m_u$ and $m_c$ 
in a reasonable range will not be able to accommodate a value
of $V_{ub}/V_{cb}=0.08$ \cite{xingpriv}.
In this regard, we are in disagreement with the analysis of 
Ref.~\cite{randhawa:99}, where much larger  values ($0.083 \sim 0.099$)
for $V_{ub}/V_{cb}$ were obtained. 
We note also that texture zeros are not invariant under change of basis.
Our result for the pair ($M_{3},M_{3}$) is valid only in the
hierarchical basis.  Otherwise, larger values of $V_{ub}/V_{cb}$ are
allowed, as in Eqs.(23-26) of Ref. \cite{branco:99}.
For these reasons, we exclude the $(M_{3},M_{3})$ pair from the ``Yes" column
in Table II. 

It is worthwhile to compare in detail the results of our two examples.
Both the pairs $(M_{8},M_{1})$ and $(M_{3},M_{3})$ have four texture zeros.
But the former yields one relation  that is not 
physically predictive in nature
(Eq.(\ref{eq:R3})), while 
the later gives rise to  two independent 
predictions  of Eqs.(\ref{eq:Vub}-\ref{eq:Vtd})  (note that Eq.~(\ref{eq:Vus3})
is not independent).

In general, for a pair of hermitian matrices with four texture zeros,
there are eight real parameters plus one or more phases, so we expect 
at most one prediction.
 This is the case for the pair $(M_{8},M_{1})$,  where we have more than one
unremovable phases and no physical prediction. 
The situation is different with $(M_{3},M_{3})$.
From Eqs.(\ref{eq:Mt'U3}-\ref{eq:Mt'D3}), 
using $a_{U}=-|b_{U}|^{2}/d_{U}$ and 
$a_{D}=-|b_{D}|^{2}/d_{D}$, the independent parameters are
$|b_{U}|$, $|b_{D}|$, $|d_{U}|$, $|d_{D}|$, $|e_{U}-e_{D}|$,
plus two overall mass scales and one physical phase.
Thus, triangularization reveals that the pair $(M_{3},M_{3})$ actually
contains one less parameter than expected, because only the
combination $(e_{U}-e_{D})$ enters.  Put another way, we could set
$e_{U}=0$ (or $e_{D}=0$) without any effect.  This means that the 
physical contents of $(M_{3},M_{3})$ are the same as the five texture zero
pair $(M_{7},M_{3})$ or $(M_{3},M_{7})$.  In addition, we can see
that the five-zero pair $(M_{8},M_{3})$ also reduces to
 the same triangular form by a LH 2-3
rotation.  Thus, we conclude that all four pairs, $(M_{3},M_{3})$,
$(M_{7},M_{3})$, $(M_{3},M_{7})$, and $(M_{8},M_{3})$, are physically
equivalent.  The analytical predictions of the last three pairs were
studied in Ref.~\cite{kmw2:99}. 

Similar analyses of  the hermitian pairs with $M_{U}^{11}\neq 0$ and/or 
$M_{D}^{11} \neq 0$ follow directly. 
The results are presented in Tables~III and IV. 
In Table III, we list the nine hierarchical  hermitian matrices 
with $(1,1) \neq 0$ 
that are used in the construction of viable hermitian pairs.
The corresponding triangular matrices are also given in the table.
The viable hermitian 4-texture-zero pairs constructed from
Tables~I and III are listed in Table~IV, together with their predictions
for quark mixing.
Whereas there is no viable pair with both $M_{U}^{11}\neq 0$
and $M_{D}^{11} \neq 0$,
nine pairs with  $M_{U}^{11}\neq 0$ and $M_{D}^{11}=0$
are found  to be compatible with data.
Of the hermitian matrices with $M_{U}^{11}=0$ and $M_{D}^{11} \neq 0$,
one pair is viable, and the 2nd pair, $(M_2, M_{12})$, 
 leads to the same prediction for $V_{ub}$ as  that of the 5th
five-texture-zero pair (see Eq.(15) in Ref.~\cite{kmw2:99}).
This pair allows two different relations 
depending on the sign of $d_U$.
While the plus-sign  relation is marginally acceptable, 
 the minus-sign relation is disfavored by data.

 Before we leave this section, we would like to point out that 
some of the disfavored hermitian pairs can give rise to nontrivial
predictions. For example, the four pairs, $(M_1, M_{17})$ and 
$(M_{i}, M_{13})$ ($i=2,4,5$), all lead to the relation:
\begin{eqnarray}
\left| V_{td} \right| & \simeq & 
\sqrt{V^2_{us} m_c/m_t + m_u/m_t}
=|V_{us}| \sqrt{m_c/m_t} \times ( 1+ {\cal O}(\lambda^2))
> 1\% \; .
\end{eqnarray}
This prediction contradicts the experimental limit
of $|V_{td}| < 1\%$, as can be derived from $B\overline{B}$ mixing
(see also Eq.~(\ref{eq:Vtdexp})).
As another example,
the hermitian pair $(M_{2}, M_{11})$ entails the relation:
\begin{eqnarray}
|V_{cb}| & \simeq & \sqrt{\frac{m_c}{m_t}}
\left(1 + \frac{m_u}{m_c V^2_{us}} \right)^{-\frac{1}{2}}
 = \sqrt{\frac{m_c}{m_t}} \times ( 1+ {\cal O}(\lambda^2)) \sim 0.06 \; ,
\end{eqnarray}
which is too large for this texture to be viable.

   A notable feature about Tables I-IV is that some texture pairs
lead to the same relation, like patterns 1, 2, 5, and 6.
It can be shown that these texture pairs are related by weak basis 
transformation,
and thus are physically equivalent.

\section{Conclusions}

Because of the many redundancies in the quark mass matrices, a lot of work
has been done in search for more restrictive patterns of matrices which are
still compatible with experiments.  In the literature, such considerations have
focused on hermitian mass matrices with a certain number of texture zeros.
It was found that the maximum allowed number of texture zeros is five.  
Amongst them one unique pattern has been identified which is most
favorable with data.
It is worthwhile to understand the situation for pairs of matrices with 
fewer texture zeros.

In this paper, we have investigated systematically hermitian 
hierarchical quark mass
matrices with four texture zeros.  By transforming the hermitian matrices
into the triangular form in the minimal parameter basis,
 and using the fact that the latter contains only
physical masses and CKM matrix elements, we can quickly rule out many of the
possible pairs of mass matrices.  For the remaining candidates, we showed that 
each pattern of texture zeros implies certain
relations between quark masses and mixing.  These relations
can be used to determine the
viability of each pair of mass matrices by comparing them with available data.
In this way we identified nineteen pairs of mass matrices which are compatible
with current data.  One pair (pattern 21) is marginal, with the final
verdict depending on more accurate experimental numbers. 
Of the nineteen viable textures, none has a parallel  structure
between the up and down mass matrices.
In particular, the popular parallel-structure pair that has been 
a focus of much attention, pattern 10, leads to a low value
of $V_{ub}/V_{cb}$, and is not favorable with present data.
We also found that, by a proper LH rotation, pattern 10
can be shown to have the
same physical contents as three (which are themselves equivalent) of
the five pairs studied earlier~\cite{kmw2:99,RRR:93}.
The asymmetry between up and down quark mass matrix textures 
 could  serve as a useful guideline
 in search for  realistic models of quark-lepton masses.

\acknowledgements 
We would like to thank Sadek Mansour for helpful discussions.
G.W. would like to thank
the National Center for Theoretical Sciences of Taiwan for its hospitality
during the final stage of this work.
S.C. is supported by the Purdue Research Foundation. 
T.K. and G.W. are supported by DOE, Grant No. DE-FG02-91ER40681 and
No. DE-FG03-96ER40969, respectively.  



  \begin{table}
 \begin{center}
 \begin{tabular}{cccc}
$M_{1}$ & $M_{2}$ & $M_{3}$ & $M_{4}$   \\     \hline   
\vspace{0.25in}   
  $\left(\begin{array}{ccc}
    0 & B' & C \\
    B'^{*} & D' & E \\
    C^{*} & E^{*} & 1 \\
    \end{array}
    \right)$ 
 &  $\left(\begin{array}{ccc}
    0 & 0 & C \\
    0 & D' & E \\
    C^{*} & E^{*} & 1 \\
    \end{array}
    \right)$
 
 & $\left(\begin{array}{ccc}
    0 & B & 0 \\
    B^{*} & D' & E \\
    0 & E^{*} & 1 \\
    \end{array}
    \right)$
 
  & $\left(\begin{array}{ccc}
    0 & B & C \\
    B^{*} & D & 0 \\
    C^{*} & 0 & 1 \\
    \end{array}
    \right)$ 
 \\ \vspace{.25in} 
   $\left(\begin{array}{ccc}
    -|C|^2 - \frac{|B|^2}{D} & B & C \\
    0 & D & E \\
    0 & 0 & 1 \\
    \end{array}
    \right)$   
  &  $\left(\begin{array}{ccc}
 A^\prime  & -CE^{*} & C \\
    0 & D & E \\
    0 & 0 & 1 \\
    \end{array}
    \right)$ 
   &  $\left(\begin{array}{ccc}
    -\frac{|B|^{2}}{D} & B & 0 \\
    0 & D & E \\
    0 & 0 & 1 \\
    \end{array}
    \right)$ 
  & $\left(\begin{array}{ccc}
    -|C|^2 - \frac{|B|^2}{D}  & B & C \\
    0 & D & 0 \\
    0 & 0 & 1 \\
    \end{array}
    \right)$  
 
 \\ \hline \hline 
      
  $M_5$ &    $M_{6}$ &  $M_{7}$   & $M_{8}$      \\  \hline
\vspace{0.25in} 
    $\left(\begin{array}{ccc}
    0 & B^\prime & C \\
    {B^\prime}^* & 0 & E \\
    C^{*} & E^{*} & 1 \\
    \end{array}
    \right)$
 &   $\left(\begin{array}{ccc}
    0 & 0 & C \\
    0 & D & 0 \\
    C^{*} & 0 & 1 \\
    \end{array}
    \right)$  
  &  $\left(\begin{array}{ccc}
    0 & B & 0 \\
    B^{*} & D & 0 \\
    0 & 0 & 1 \\
    \end{array}
    \right)$
 
   & $\left(\begin{array}{ccc}
    0 & B & 0 \\
    B^{*} & 0 & E \\
    0 & E^{*} & 1 \\
    \end{array}
    \right)$ 
  \\ \vspace{0.25in} 
   $\left(\begin{array}{ccc}
    \frac{|B|^{2}}{|E|^{2}} - |C|^2  & B & C \\
    0 & -|E|^{2} & E \\
    0 & 0 & 1 \\
    \end{array}
    \right)$
 &  $\left(\begin{array}{ccc}
    -|C|^{2} & 0 & C \\
    0 & D & 0 \\
    0 & 0 & 1 \\
    \end{array}
    \right)$
    
  & $\left(\begin{array}{ccc}
    -\frac{|B|^{2}}{D} & B & 0 \\
    0 & D & 0 \\
    0 & 0 & 1 \\
    \end{array}
    \right)$ 

   &  $\left(\begin{array}{ccc}
    \frac{|B|^{2}}{|E|^{2}} & B & 0 \\
    0 & -|E|^{2} & E \\
    0 & 0 & 1 \\
    \end{array}
    \right)$
 
  \\  
    \end{tabular}
 \caption{The eight hierarchical hermitian matrices with (1,1)=0.  
The corresponding triangular forms are also listed.  
Matrices in this table can be paired up to form quark mass matrices 
with 4 texture zeros.  Here  
$A'\equiv -|C|^{2}(1+\frac{|E|^{2}}{D})$, 
      $B'\equiv B+CE^{*}$, and
       $D'\equiv D+|E|^{2}$.
 } 
   \end{center}
 \end{table}

\begin{table}
 \begin{center}
 \begin{tabular}{cclc}
    Pattern  &  $(M^{11}_{U}=0,M^{11}_{D}=0)$   &  Relation  &  Viable   \\
\hline
      1  &   $(M_{1},M_{7})$   
 & R1:  $|V_{ub}-K \cdot V_{ts}^{*}|^2 \simeq \left| 
\frac{m_{u}}{m_{t}} \pm  \frac{m_{c}}{m_{t}}|K^{2}| \right|$ 
 &  Yes \\  
    2  &   $(M_{2},M_{3})$
 & R1:  $|V_{ub}-K \cdot V_{ts}^{*}|^2 \simeq \left| 
\frac{m_{u}}{m_{t}} \pm  \frac{m_{c}}{m_{t}}|K^{2}| \right|$ 
 &  Yes \\
   3  &   $(M_{2},M_{4})$
 & R2:  $\left|\frac{V_{us}}{V_{cs}}\right| \simeq 
   \left| \sqrt{\frac{m_d}{m_s}} - e^{i\delta} \sqrt{\frac{m_u}{m_c}} 
 \sqrt{\frac{V_{cb}^2}{(m_c/m_t) \pm V_{cb}^2}} \right|$ 
 & Yes  \\
  
4  &   $(M_{3},M_{4})$
 & R3: 
$\left|\frac{V_{us}}{V_{cs}} \right| \simeq
\left|\sqrt{\frac{m_d}{m_s}} - e^{i\delta} \sqrt{\frac{m_u}{m_c}} \right|$
  & Yes \\  
    5  &   $(M_{4},M_{3})$
 & R1:  $|V_{ub}-K \cdot V_{ts}^{*}|^2 \simeq \left| 
\frac{m_{u}}{m_{t}} \pm  \frac{m_{c}}{m_{t}}|K^{2}| \right|$ 
 &  Yes \\
   6  &   $(M_{5},M_{3})$
 & R1:  $|V_{ub}-K \cdot V_{ts}^{*}|^2 \simeq \left| 
\frac{m_{u}}{m_{t}} \pm  \frac{m_{c}}{m_{t}}|K^{2}| \right|$ 
 & Yes  \\
   7  &   $(M_{6},M_{1})$
 & R4: $|V_{us}| \simeq \sqrt{\frac{m_d}{m_s}}$
&  Yes \\
 8  &   $(M_{7},M_{1})$
 & R3: 
$\left|\frac{V_{us}}{V_{cs}} \right| \simeq
\left|\sqrt{\frac{m_d}{m_s}} - e^{i\delta} \sqrt{\frac{m_u}{m_c}} \right|$
&  Yes \\   
 9   &   $(M_{8},M_{1})$
 & R3: 
$\left|\frac{V_{us}}{V_{cs}} \right| \simeq
\left|\sqrt{\frac{m_d}{m_s}} - e^{i\delta} \sqrt{\frac{m_u}{m_c}} \right|$
&  Yes \\
\hline
$10$  &   $(M_{3},M_{3})$
 & R5:  $\left|\frac{V_{td}}{V_{ts}}\right|  \simeq \sqrt{\frac{m_{d}}{m_{s}}}$,
  R6:  $\left|\frac{V_{ub}}{V_{cb}}\right|  \simeq \sqrt{\frac{m_{u}}{m_{c}}}$,
   and  R3 
    & No \\
    \end{tabular}
   \caption{The ten possible candidates for hermitian, hierarchical 
quark mass matrices with vanishing (1,1) elements.
Also listed are the testable relations derived from each pair. 
In writing relation R1, we have used for simplicity in presentation
$V_{tb} \simeq V_{cs} \simeq V_{ud} \simeq 1$, and 
$K \equiv e^{i\delta} \sqrt{\frac{m_{d}}{m_{s}}}-\frac{V_{us}}{V_{cs}}$.
    Except for pattern 7, all other relations require properly
   chosen phases to be viable. 
 The nine viable pairs lead to 4 different relations, 
three of which are similar: R2, R3, and R4.
Relation R6 predicts a too small $V_{ub}/V_{cb}$,
 which makes pattern 10 disfavored.
 } 
   \end{center}
 \end{table}

  \begin{table}
 \begin{center}
 \begin{tabular}{ccccc}
$M_{9}$ & $M_{10}$ & $M_{11}$ & $M_{12}$ & $M_{13}$  \\   \hline
\vspace{0.25in}
$\left(\begin{array}{ccc}
    A^{\prime\prime} & 0 & C \\
    0 & D' & E \\
    C^{*} & E^{*} & 1 \\
    \end{array}
    \right)$
 &  $\left(\begin{array}{ccc}
    A + \frac{|B|^2}{D} & B & 0 \\
    B^{*} & D' & E \\
    0 & E^{*} & 1 \\
    \end{array}
    \right)$

  & $\left(\begin{array}{ccc}
    A+|C|^2 & 0 & C \\
    0 & D & 0 \\
    C^{*} & 0 & 1 \\
    \end{array}
    \right)$

  &  $\left(\begin{array}{ccc}
    A +\frac{|B|^2}{D} & B & 0 \\
    B^{*} & D & 0 \\
    0 & 0 & 1 \\
    \end{array}
    \right)$

 &  $\left(\begin{array}{ccc}
    A & 0 & 0 \\
    0 & D' & E \\
    0 & E^{*} & 1 \\
    \end{array}
    \right)$
\\
    $\left(\begin{array}{ccc}
    A &-CE^{*}  & C \\
    0 & D & E \\
    0 & 0 & 1 \\
    \end{array}
    \right)$

&  $\left(\begin{array}{ccc}
    A & B & 0 \\
    0 & D & E \\
    0 & 0 & 1 \\
    \end{array}
    \right)$

  & $\left(\begin{array}{ccc}
    A & 0 & C \\
    0 & D & 0 \\
    0 & 0 & 1 \\
    \end{array}
    \right)$
  &  $\left(\begin{array}{ccc}
    A & B & 0 \\
    0 & D & 0 \\
    0 & 0 & 1 \\
    \end{array}
    \right)$

&  $\left(\begin{array}{ccc}
    A & 0 & 0 \\
    0 & D & E \\
    0 & 0 & 1 \\
    \end{array}
    \right)$
 \\ \hline \hline
$M_{14}$ & $M_{15}$ & $M_{16}$ & $M_{17}$ &      \\     \hline
\vspace{0.25in}

  $\left(\begin{array}{ccc}
    A-\frac{|B|^2}{|E|^2} & B & 0 \\
    B^{*} & 0 & E \\
    0 & E^{*} & 1 \\
    \end{array}
    \right)$

 & $\left(\begin{array}{ccc}
    A & 0 & C \\
    0 & 0 & E \\
    C^{*} & E^{*} & 1 \\
    \end{array}
    \right)$
  & $\left(\begin{array}{ccc}
    A & 0 & 0 \\
    0 & 0 & E \\
    0 & E^{*} & 1 \\
    \end{array}
    \right)$

&  $\left(\begin{array}{ccc}
    A & 0 & 0 \\
   0 & D & 0 \\
    0 & 0 & 1 \\
    \end{array}
    \right)$
& 
\\

 $\left(\begin{array}{ccc}
    A & B & 0 \\
    0 & -|E|^{2} & E \\
    0 & 0 & 1 \\
    \end{array}
    \right)$

& $\left(\begin{array}{ccc}
    A & -CE^{*} & C \\
    0 & -|E|^{2} & E \\
    0 & 0 & 1 \\
    \end{array}
    \right)$
 &  $\left(\begin{array}{ccc}
    A & 0 & 0 \\
    0 & -|E|^{2} & E \\
    0 & 0 & 1 \\
    \end{array}
    \right)$

& $\left(\begin{array}{ccc}
    A & 0 & 0 \\
    0 & D & 0 \\
    0 & 0 & 1 \\
    \end{array}
    \right)$
&
\end{tabular}
\caption{Nine possible candidates for $M_U$ or $M_D$ with nonzero $(1,1)$
element.
 Both hermitian and triangular forms are given in the hierarchical basis.
 Here 
$D'\equiv D+|E|^{2}$, and 
$A^{\prime\prime} \equiv A+\frac{|CE|^{2}}{D}+|C|^{2}$. } 
\end{center}
\end{table}


\begin{table}
 \begin{center}
 \begin{tabular}{cclc}
    Pattern &  $(M^{11}_{U}\neq 0, M^{11}_{D}=0)$   &  Relation  & Viable  \\
\hline
     11  &   $(M_{9},M_{7})$
  & R7: $|V_{cb}| \simeq \sqrt{\frac{m_c}{m_t}} 
 \left| 1 - \frac{V_{ub}}{V_{ts}^* K} \right|^{-\frac{1}{2}}$
 & Yes  \\
    12  &   $(M_{10},M_{7})$
 & R5:  $\left|\frac{V_{td}}{V_{ts}}\right| \simeq \sqrt{ \frac{m_{d}}{m_{s}}}$
 & Yes  \\
  13  &   $(M_{11},M_{3})$
 & R4:  $ |V_{us}|  \simeq \sqrt{\frac{m_{d}}{m_{s}}}$
  &  Yes \\
   14  &   $(M_{12},M_{3})$
 & R5:  $\left|\frac{V_{td}}{V_{ts}}\right| \simeq \sqrt{ \frac{m_{d}}{m_{s}}}$
&  Yes \\
 15  &   $(M_{13},M_{4})$
 & R4:  $ |V_{us}|  \simeq \sqrt{\frac{m_{d}}{m_{s}}}$
&  Yes \\
 16  &   $(M_{14},M_{3})$
 & R5:  $\left|\frac{V_{td}}{V_{ts}}\right| \simeq \sqrt{ \frac{m_{d}}{m_{s}}}$
&  Yes \\

17  &   $(M_{15},M_{3})$
  & R8: $|V_{cb}| \simeq \sqrt{\frac{m_c}{m_t}} 
 \left| 1 - \frac{V_{ub}}{V_{ts}^* K} \right|^{-1}$
&  Yes \\
 
 18  &   $(M_{16},M_{1})$
 & R4:  $ |V_{us}|  \simeq \sqrt{\frac{m_{d}}{m_{s}}}$
&  Yes \\

19  &   $(M_{17},M_{1})$
 & R4:  $ |V_{us}|  \simeq \sqrt{\frac{m_{d}}{m_{s}}}$
& Yes  \\
\hline
    Pattern  &  $(M^{11}_{U}=0,M^{11}_{D}\neq 0)$   &  Relation  & Viable \\
\hline
     20  &   $(M_{6},M_{10})$
     & R9: $ |V_{ub}|  \simeq \sqrt{\frac{m_{u}}{m_{t}}}$
& Yes  \\
 21  &   $(M_{2},M_{12})$ 
   & R10: $|V_{ub}| \simeq \sqrt{\frac{m_{u}}{m_{t}}}
  \Gamma_{\pm}^{-\frac{1}{2}}$
& ${\#}$  \\
 22 & $(M_{3},M_{12})$  &   
R6: $\left| \frac{V_{ub}}{V_{cb}} \right| \simeq \sqrt{\frac{m_{u}}{m_{c}}}$
& No \\
 23 & $(M_{7},M_{10})$  &   
R6: $\left| \frac{V_{ub}}{V_{cb}} \right| \simeq \sqrt{\frac{m_{u}}{m_{c}}}$
& No \\
 24  & $(M_{8},M_{10})$  &   
R6: $\left| \frac{V_{ub}}{V_{cb}} \right| \simeq \sqrt{\frac{m_{u}}{m_{c}}}$
& No \\
    \end{tabular}
   \caption{Possible hermitian pairs and their corresponding relations for
   ($M_{U}^{11} \neq 0$,  $M_{D}^{11}=0$) and 
  ($M_{U}^{11} = 0$,  $M_{D}^{11} \neq 0$).  
Here $\Gamma_{\pm}^{-1} \equiv 1 \pm V^2_{cb} m_t/m_c \simeq 1 \pm 0.4$ at the
scale $m_Z$.
In writing relations R7 and R8, we have used for simplicity
$V_{tb} \simeq V_{cs} \simeq V_{ud} \simeq 1 $, and 
$K \equiv e^{i\delta} \sqrt{\frac{m_{d}}{m_{s}}}-\frac{V_{us}}{V_{cs}}$.
The particular pair (pattern 21) that leads to predictions marginally 
acceptable by data is indicated by \#.
Note that no viable 4-texture-zero pair exists with $M_{U}^{11} \neq 0$ and 
$M_{D}^{11} \neq 0$.
}
   \end{center}
 \end{table}

\end{document}